# On the Combined Effect of Directional Antennas and Imperfect Spectrum Sensing upon Ergodic Capacity of Cognitive Radio Systems

Hassan Yazdani, Azadeh Vosoughi, *Senior Member, IEEE*
University of Central Florida
E-mail: `h.yazdani@knights.ucf.edu, azadeh@ucf.edu`

***Abstract*—We consider a cognitive radio system, consisting of a primary transmitter ($\mathrm{PU_{tx}}$), a primary receiver ($\mathrm{PU_{rx}}$), a secondary transmitter ($\mathrm{SU_{tx}}$), and a secondary receiver ($\mathrm{SU_{rx}}$). The secondary users (SUs) are equipped with steerable directional antennas. We assume the SUs and primary users (PUs) coexist and the $\mathrm{SU_{tx}}$ knows the geometry of network. We find the ergodic capacity of the channel between $\mathrm{SU_{tx}}$ and $\mathrm{SU_{rx}}$, and study how spectrum sensing errors affect the capacity. In our system, the $\mathrm{SU_{tx}}$ first senses the spectrum and then transmits data at two power levels, according to the result of sensing. The optimal $\mathrm{SU_{tx}}$ transmit power levels and the optimal directions of $\mathrm{SU_{tx}}$ transmit antenna and $\mathrm{SU_{rx}}$ receive antenna are obtained by maximizing the ergodic capacity, subject to average transmit power and average interference power constraints. To study the effect of fading channel, we considered three scenarios: 1) when $\mathrm{SU_{tx}}$ knows fading channels between $\mathrm{SU_{tx}}$ and $\mathrm{PU_{rx}}$, $\mathrm{PU_{tx}}$ and $\mathrm{SU_{rx}}$, $\mathrm{SU_{tx}}$ and $\mathrm{SU_{rx}}$, 2) when $\mathrm{SU_{tx}}$ knows only the channel between $\mathrm{SU_{tx}}$ and $\mathrm{SU_{rx}}$, and statistics of the other two channels, and, 3) when $\mathrm{SU_{tx}}$ only knows the statistics of these three fading channels. For each scenario, we explore the optimal $\mathrm{SU_{tx}}$ transmit power levels and the optimal directions of $\mathrm{SU_{tx}}$ and $\mathrm{SU_{rx}}$ antennas, such that the ergodic capacity is maximized, while the power constraints are satisfied.**

## I. Introduction

Cognitive radio (CR) systems can alleviate spectrum scarcity problem by allowing an unlicensed user to access licensed bands under the condition that its imposed interference on the licensed users are limited [1]. Optimizing the transmission strategies of secondary users (SUs) in the presence of a primary user (PU) has attracted much research interests in industry and academia [2]–[10], where most of these works assume the SUs are equipped with *omni-directional* antennas and the result of spectrum sensing is *perfect*. However, spectrum sensing methods are prone to errors and their false alarm and detection probabilities should be incorporated in the design and performance analysis. Different from the bulk of the literature, in this paper we assume the SUs and PUs can coexist, the $\mathrm{SU_{tx}}$ knows the geometry of network. Also, SUs are equipped with *steerable directional* antennas and can use *spatial spectrum holes* [11]–[13] to increase spectrum utilization.

In this work, the SU transmitter ($\mathrm{SU_{tx}}$) first senses the spectrum and then adapts its transmit power, according to the result of spectrum sensing, i.e., $\mathrm{SU_{tx}}$ transmits signal to secondary receiver ($\mathrm{SU_{rx}}$) with power levels $P_0$ and $P_1$ when spectrum is sensed idle and busy, respectively. To study the effect of fading channels, we consider three scenarios: 1) when $\mathrm{SU_{tx}}$ has channel state information (CSI) of links between $\mathrm{SU_{tx}}$ and $\mathrm{PU_{rx}}$, $\mathrm{PU_{tx}}$ and $\mathrm{SU_{rx}}$, $\mathrm{SU_{tx}}$ and $\mathrm{SU_{rx}}$, 2) when $\mathrm{SU_{tx}}$ knows only the CSI of link between $\mathrm{SU_{tx}}$ and $\mathrm{SU_{rx}}$, and the statistics of the other two links, and, 3) when $\mathrm{SU_{tx}}$ only knows the statistics of these three fading channels. For each scenario, we establish the ergodic capacity of the channel between $\mathrm{SU_{tx}}$ and $\mathrm{SU_{rx}}$, when spectrum sensing is *imperfect* and find the optimal directions of $\mathrm{SU_{tx}}$ and $\mathrm{SU_{rx}}$ antennas and optimal $\mathrm{SU_{tx}}$ power levels such that the ergodic capacity is maximized, subject to average transmit power and average interference power constraints.

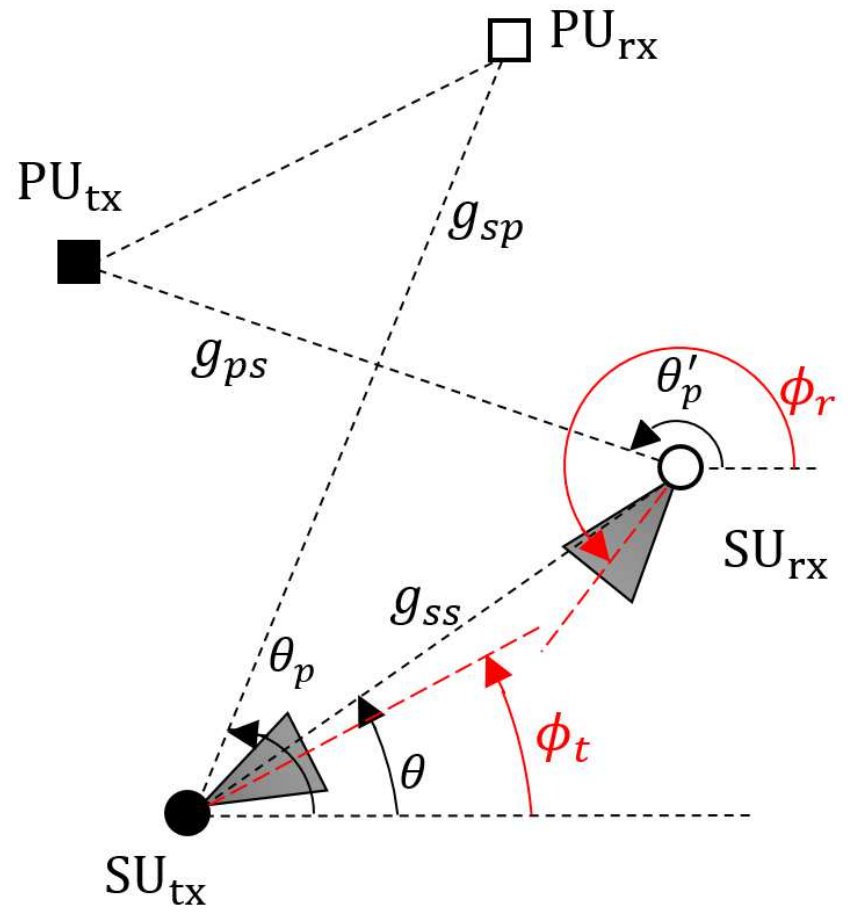

Fig. 1: Our cognitive radio system with directional antennas.

## II. System Model

Our CR system model is shown in Fig. 1. The SUs are equipped with steerable directional antennas. The orientation of $\mathrm{PU_{rx}}$ and $\mathrm{SU_{rx}}$ with respect to $\mathrm{SU_{tx}}$ are denoted by $\theta_p$ and $\theta$, receptively, and the orientation of $\mathrm{PU_{tx}}$ with respect to $\mathrm{SU_{rx}}$ is denoted by $\theta'_p$. The boresight of $\mathrm{SU_{tx}}$ and $\mathrm{SU_{rx}}$ antennas in their local coordination are denoted by $\phi_t$ and $\phi_r$, respectively. We assume $\theta_p$, $\theta$ and $\theta'_p$ are known or can be estimated [14]. The antenna gain is given by $A(\phi)=A_1+A_0\ \exp\left(-B\left(\frac{\phi}{\phi_{3\text{dB}}}\right)^2\right)$ where $B=\ln(2)$, $\phi_{3\text{dB}}$ is the half-power beam-width, $A_1$ and $A_0$ are two constant parameters [12], [13]. Let $d_{ps}$, $d_{sp}$ and $d_{ss}$ be the distances between $\mathrm{PU_{tx}}$ and $\mathrm{SU_{rx}}$, $\mathrm{PU_{rx}}$ and $\mathrm{SU_{tx}}$, and $\mathrm{SU_{tx}}$ and $\mathrm{SU_{rx}}$, respectively.

Spectrum sensing at the $\text{SU}_{\text{tx}}$ can be formulated as a binary hypothesis testing problem in which $\mathcal{H}_0$ and $\mathcal{H}_1$ with prior probabilities $\pi_0$ and $\pi_1 = 1 - \pi_0$ denote the spectrum is truly idle and truly busy, respectively. When the spectrum is truly busy, the average transmit power of $\text{PU}_{\text{tx}}$ is $P_p$ and we assume $\text{SU}_{\text{tx}}$ knows $P_p$. Let $\hat{\mathcal{H}}_1$ and $\hat{\mathcal{H}}_0$ with probabilities $\hat{\pi}_0 = \Pr\{\hat{\mathcal{H}}_0\}$ and $\hat{\pi}_1 = \Pr\{\hat{\mathcal{H}}_1\}$ denote that the result of spectrum sensing is busy and idle, respectively. When the spectrum is sensed idle and busy, $\text{SU}_{\text{tx}}$ uses two power levels $P_0$ and $P_1$, respectively to transmit signal to $\text{SU}_{\text{rx}}$. The accuracy of spectrum sensing method is characterized by false alarm probability $P_f = \Pr\{\hat{\mathcal{H}}_1|\mathcal{H}_0\}$ and detection probability $P_d = \Pr\{\hat{\mathcal{H}}_1|\mathcal{H}_1\}$. We assume $\pi_0$, $P_d$, $P_f$ are known.

The fading from $\text{SU}_{\text{tx}}$ to $\text{SU}_{\text{rx}}$, and $\text{PU}_{\text{tx}}$ to $\text{SU}_{\text{rx}}$ are denoted by $g_{ss}$ and $g_{ps}$, respectively, and $g_{sp}$ is the fading from $\text{SU}_{\text{tx}}$ to $\text{PU}_{\text{rx}}$. We assume $g_{ss}$, $g_{ps}$ and $g_{sp}$ are three independent exponentially distributed random variables with mean $\gamma_{ss}$, $\gamma_{ps}$ and $\gamma_{sp}$, respectively. The path-loss is $L = (d_0/d)^{\nu}$, where $d_0$ is the reference distance, $d$ is the distance between users, and $\nu$ is the path loss exponent. Our goal is to find the ergodic capacity of the channel between $\text{SU}_{\text{tx}}$ and $\text{SU}_{\text{rx}}$ and explore the optimal SU transmit power levels and the optimal directions of $\text{SU}_{\text{tx}}$ and $\text{SU}_{\text{rx}}$ antennas, such that this capacity maximized, subject to average transmit power and average interference power constraints.

## III. Constrained Ergodic Capacity Maximization

When spectrum sensing is imperfect, depending on the true status of the PU and the spectrum sensing result, the ergodic capacity can be written as $C = \mathbb{E}_{\boldsymbol{g}}\Big\{ \sum_{i=0}^{1} \big(\alpha_i \, c_{0,i} + \beta_i \, c_{1,i}\big)\Big\}$, where $\mathbb{E}_{\boldsymbol{g}}\{.\}$ is the expectation operator over random fading coefficients $\boldsymbol{g} = \big(g_{ss}, g_{sp}, g_{ps}\big)$ and $c_{i,j}$ is instantaneous capacity corresponding to $\mathcal{H}_i$ and $\hat{\mathcal{H}}_j$ with probability $\alpha_i = \Pr\{\mathcal{H}_0, \hat{\mathcal{H}}_i\}$ and $\beta_i = \Pr\{\mathcal{H}_1, \hat{\mathcal{H}}_i\}$ for $i \in \{0, 1\}$, given as

$$c_{0,i} = \log_2\left(1 + \frac{g_{ss}L_{ss}\; G(\theta, \phi_t, \phi_r)P_i(\boldsymbol{g})}{\sigma_n^2}\right) \quad (1)$$

$$c_{1,i} = \log_2\left(1 + \frac{g_{ss}L_{ss}\; G(\theta, \phi_t, \phi_r)P_i(\boldsymbol{g})}{\sigma_n^2 + P_p\; g_{ps}L_{ps}\; A(\phi_r - \theta_p')}\right). \quad (2)$$

In (1) and (2), $G(\theta, \phi_t, \phi_r) = A(\phi_t - \theta)A(\phi_r - \pi - \theta)$ is the product of $\text{SU}_{\text{tx}}$ and $\text{SU}_{\text{rx}}$ antennas' gain and $\sigma_n^2$ is the variance of additive zero-mean Gaussian noise at $\text{SU}_{\text{rx}}$. It is easy to verify

$$\alpha_0 = \pi_0(1 - P_f), \quad \alpha_1 = \pi_0 P_f,$$
$$\beta_0 = \pi_1(1 - P_d), \quad \beta_1 = \pi_1 P_d.$$

Note that the optimal antenna directions $\phi_t$ and $\phi_r$ are expected to be functions of fading $\boldsymbol{g}$ and for simplicity, we dropped parameter $\boldsymbol{g}$. Also, for simplicity of presentation, we drop the parameters $\theta$, $\phi_t$ and $\phi_r$ from $G(\theta, \phi_t, \phi_r)$ and define $a = g_{ss}L_{ss}G$ and $\sigma_p^2 = P_p\; g_{ps}L_{ps}A(\phi_r - \theta_p')$. The term $\sigma_p^2$ captures the interference on $\text{SU}_{\text{rx}}$ due to PU activities. Then, we can rewrite (1) and (2) as $c_{0,i} = \log_2\left(1 + \frac{aP_i(\boldsymbol{g})}{\sigma_n^2}\right)$ and $c_{1,i} = \log_2\left(1 + \frac{aP_i(\boldsymbol{g})}{\sigma_n^2 + \sigma_p^2}\right)$, respectively.

Let $\bar{I}_{\text{av}}$ indicate the maximum allowed interference power of $\text{PU}_{\text{rx}}$ and $\bar{P}_{\text{av}}$ denote the maximum allowed average transmit power of $\text{SU}_{\text{tx}}$. To satisfy the average interference power constraint, we have

$$\mathbb{E}_{\boldsymbol{g}}\left\{ \Big(\beta_0 P_0(\boldsymbol{g}) + \beta_1 P_1(\boldsymbol{g})\Big)\; g_{sp}L_{sp}\; A(\phi_t - \theta_p)\right\} \leq \bar{I}_{\text{av}}. \quad (3)$$

By defining $b_i = \beta_i g_{sp} L_{sp} A(\phi_t - \theta_p)$, (3) can be written as

$$\mathbb{E}_{\boldsymbol{g}}\Big\{b_0 P_0(\boldsymbol{g}) + b_1 P_1(\boldsymbol{g})\Big\} \leq \bar{I}_{\text{av}}. \quad (4)$$

In (4), $b_0P_0(\boldsymbol{g})$ and $b_1P_1(\boldsymbol{g})$ denote the imposed interference to $\text{PU}_{\text{rx}}$ from $\text{SU}_{\text{tx}}$ when channel is sensed idle and busy, respectively. To satisfy the average transmit power constraint, we have

$$\mathbb{E}_{\boldsymbol{g}}\{\hat{\pi}_0 P_0(\boldsymbol{g}) + \hat{\pi}_1 P_1(\boldsymbol{g})\} \leq \bar{P}_{\text{av}}. \quad (5)$$

The problem we consider is maximizing the ergodic capacity $C$ over $P_0(\boldsymbol{g}), P_1(\boldsymbol{g}), \phi_t$ and $\phi_r$ subject to constraints (4) and (5). The expression $C$ is concave with respect to $P_0(\boldsymbol{g})$, $P_1(\boldsymbol{g})$ and $\phi_r$. However, it is not concave with respect to $\phi_t$. The optimal $\phi_t$ can be obtained using one-dimensional search, i.e., we consider an initial value for $\phi_t$ and find $P_0(\boldsymbol{g})$, $P_1(\boldsymbol{g})$ and $\phi_r$. Then, we find the value of $\phi_t$ which maximizes $C$. Given $\phi_t$, we can solve this problem using the Lagrange multipliers method to find $P_0(\boldsymbol{g})$, $P_1(\boldsymbol{g})$ and $\phi_r$. The Lagrangian is given as

$$L = -\mathbb{E}_{\boldsymbol{g}}\Big\{ \sum_{i=0}^{1} \big(\alpha_i\, c_{0,i} + \beta_i\, c_{1,i}\big)\Big\} + \lambda\Big(\mathbb{E}_{\boldsymbol{g}}\Big\{\hat{\pi}_0 P_0(\boldsymbol{g}) + \hat{\pi}_1 P_1(\boldsymbol{g})\Big\} - \bar{P}_{\text{av}}\Big) + \mu\Big(\mathbb{E}_{\boldsymbol{g}}\Big\{b_0 P_0(\boldsymbol{g}) + b_1 P_1(\boldsymbol{g})\Big\} - \bar{I}_{\text{av}}\Big) \quad (6)$$

where $\lambda$ and $\mu$ are nonnegative Lagrange multipliers. In the following subsections, we address this constrained maximization problem when 1) $\text{SU}_{\text{tx}}$ knows perfect CSI of $\boldsymbol{g}$, 2) when $\text{SU}_{\text{tx}}$ knows only $g_{ss}$, and statistics of $g_{ps}$ and $g_{sp}$, 3) when $\text{SU}_{\text{tx}}$ only knows the statistics of $\boldsymbol{g}$.

### A. Perfect CSI for Three Fading Channels

In the first scenario, we assume $\text{SU}_{\text{tx}}$ has perfect knowledge of $g_{ss}$, $g_{ps}$ and $g_{sp}$ and it maximizes the capacity for each realization of fading coefficients. Taking the derivative of Lagrangian in (6) with respect to $P_i(\boldsymbol{g})$ and equaling it to zero gives

$$\frac{\partial L}{\partial P_i(\boldsymbol{g})} = \frac{-a}{\sigma_n^2 \ln(2)} w_i(x, y) + \lambda \hat{\pi}_i + \mu b_i = 0 \quad (7)$$

where $y \triangleq \sigma_n^2/\sigma_p^2$, $x_i \triangleq \sigma_n^2/aP_i(\boldsymbol{g})$ and

$$w_i(x, y) = x\left(\frac{\alpha_i}{x+1} + \frac{\beta_i y}{xy + x + y}\right).$$

Also, $x^{-1}$ and $y^{-1}$ are the received signal-to-noise-ratio (SNR) and interference-to-noise-ratio (INR) at $\text{SU}_{\text{rx}}$. By solving (7), the optimal transmit power levels can be written as

$$P_i(\boldsymbol{g}) = \left[\frac{F_i + \sqrt{\Delta_i}}{2}\right]^+ \qquad \text{for } i = 0, 1 \quad (8)$$

where $[x]^+$ denotes $\max(x, 0)$ and

$$F_i = \frac{\hat{\pi}_i}{\ln(2)\,(\lambda \hat{\pi}_i + \mu b_i)} - \frac{2\sigma_n^2 + \sigma_p^2}{a}$$

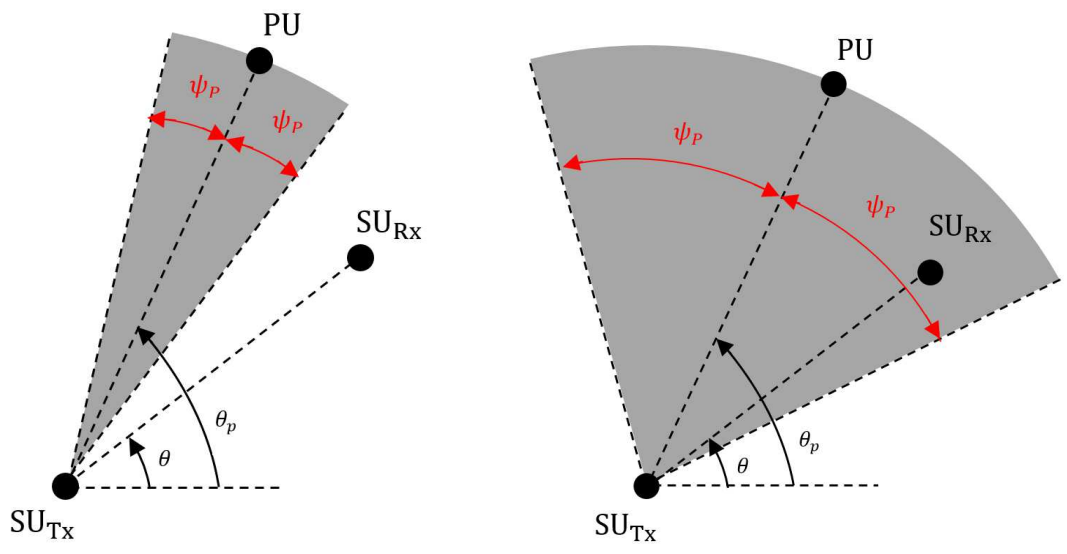

(a) when $|\theta_p - \theta| > \psi_p$ (b) when $|\theta_p - \theta| < \psi_p$

Fig. 2: Illustration of $\phi_t^{\text{opt}}$ for $0 < Z \leq 1$

$$\Delta_i = F_i^2 - \frac{4}{a}\left(\frac{\sigma_n^2(\sigma_n^2+\sigma_p^2)}{a} - \frac{\hat{\pi}_i\sigma_n^2+\beta_i\sigma_p^2}{\ln(2)\left(\lambda\hat{\pi}_i+\mu b_i\right)}\right).$$

The Lagrange multipliers $\lambda$ and $\mu$ can be updated using subgradient method as follows [5]

$$\lambda^{(n+1)} = \left[\lambda^{(n)} + t_0\Big(\mathbb{E}_{\boldsymbol{g}}\big\{\hat{\pi}_0 P_0(\boldsymbol{g}) + \hat{\pi}_1 P_1(\boldsymbol{g})\big\} - \bar{P}_{\text{av}}\Big)\right]^+ \quad (9a)$$

$$\mu^{(n+1)} = \left[\mu^{(n)} + t_0\Big(\mathbb{E}_{\boldsymbol{g}}\big\{b_0 P_0(\boldsymbol{g}) + b_1 P_1(\boldsymbol{g})\big\} - \bar{I}_{\text{av}}\Big)\right]^+ \quad (9b)$$

where $t_0$ is the step size and $\lambda$ and $\mu$ converge when for a small number $\delta$ we get

$$\lambda^{(n)}\Big(\mathbb{E}_{\boldsymbol{g}}\big\{\hat{\pi}_0 P_0(\boldsymbol{g}) + \hat{\pi}_1 P_1(\boldsymbol{g})\big\} - \bar{P}_{\text{av}}\Big) \leq \delta \quad (10a)$$

$$\mu^{(n)}\Big(\mathbb{E}_{\boldsymbol{g}}\big\{b_0 P_0(\boldsymbol{g}) + b_1 P_1(\boldsymbol{g})\big\} - \bar{I}_{\text{av}}\Big) \leq \delta. \quad (10b)$$

The optimal $\phi_r$ can be obtained by solving $\partial L/\partial \phi_r = 0$. There is no closed form solution for $\phi_r^{\text{opt}}$, but, one can verify that when transmit power of $\text{PU}_{\text{tx}}$ is zero ($P_p = 0$), $\phi_r^{\text{opt}} = \pi + \theta$. We can reduce the computational complexity of one-dimensional search for finding $\phi_t^{\text{opt}}$ by finding a narrower interval to which $\phi_t^{\text{opt}}$ belongs to [13]. We define

$$Z = \frac{\bar{I}_{\text{av}}}{\pi_1 g_{sp} A_0 \bar{P}_{\text{av}}} - \frac{A_1}{A_0}. \quad (11)$$

If $Z > 1$, it means that $\text{PU}_{\text{rx}}$ can tolerate an interference power that is larger than the interference power imposed by $\text{SU}_{\text{tx}}$, constraint (4) is loose and $\phi_t^{\text{opt}} = \theta$. When $0 < Z \leq 1$, we define $\psi_p = \phi_{3\text{dB}}\sqrt{\frac{-1}{B}\ln(Z)}$ and consider two cases. When $|\theta_p - \theta| > \psi_p$, $\phi_t^{\text{opt}}$ has to lie outside the shaded area shown in Fig. 2a. Since the unshaded area in Fig. 2a includes the line of sight (LOS) between $\text{SU}_{\text{tx}}$ and $\text{SU}_{\text{rx}}$, $\phi_t^{\text{opt}} = \theta$. When $|\theta_p - \theta| < \psi_p$, which is shown in Fig. 2b, $\phi_t^{\text{opt}}$ lies in the

$$\begin{cases} \phi_t^{\text{opt}} \in [\theta_p - \psi_p, \theta], & \text{if } \theta_p > \theta \\ \phi_t^{\text{opt}} \in [\theta, \theta_p + \psi_p], & \text{if } \theta_p < \theta. \end{cases}$$

If $Z \leq 0$, we cannot find a narrower interval. Algorithm 1 summarizes our proposed approach to find the optimal solutions $\phi_t^{\text{opt}}$, $\phi_r^{\text{opt}}$, $P_0^{\text{opt}}$ and $P_1^{\text{opt}}$.

**Algorithm 1:** Optimization Algorithm

```
k ← 0
φ_r^(0) = π + θ
repeat
    λ^(0) = λ_init, μ^(0) = μ_init
    n ← 0
    repeat
        calculate P_0^(k) and P_1^(k) using (8).
        update λ and μ using (9).
        n ← n + 1
    until (10) is satisfied;
    solve ∂L/∂φ_r = 0 and update φ_r^(k+1).
    k ← k + 1
until the differences of φ_r^(k), P_0^(k) and P_1^(k) in two
  consecutive iterations is less than some pre-determined
  values;
φ_t^opt = argmax {C} using bisection search
P_i^opt = [P_i]_{φ_t = φ_t^opt}
φ_r^opt = [φ_r]_{φ_t = φ_t^opt}
```

### B. Perfect CSI for $g_{ss}$ and Statistical CSI for Other Channels

For the second scenario, we assume that SUs cannot cooperate with PUs and as a result, $\text{SU}_{\text{tx}}$ and $\text{SU}_{\text{rx}}$ cannot estimate the fading coefficients $g_{sp}$ and $g_{ps}$, respectively and they only know the statistics of fading coefficients $g_{sp}$ and $g_{ps}$. On the other hand, we assume that $\text{SU}_{\text{tx}}$ has perfect knowledge of fading coefficient $g_{ss}$. Therefore, at first we take expectation with respect to $g_{sp}$ and $g_{ps}$ in ergodic capacity expression and then maximize the capacity. In this case the optimal transmit power levels and the optimal antenna directions are functions of $g_{ss}$. The instantaneous capacity $c_{0,i}$ is independent of $g_{sp}$ and $g_{ps}$ and $\mathbb{E}_{g_{ps},g_{sp}}\{c_{0,i}\} = c_{0,i}$. The expectation of $c_{1,i}$ can be written as

$$\mathbb{E}_{g_{ps},g_{sp}}\{c_{1,i}\} = \mathbb{E}_{g_{ps}}\left\{\log_2\left(1+\frac{g_{ss}L_{ss}GP_i(g_{ss})}{\sigma_n^2+P_p\, g_{ps}L_{ps}A(\phi_r-\theta_p')}\right)\right\}$$
$$= \frac{1}{\ln(2)}\left[\ln\left(1+\frac{1}{x_i}\right) + T(\bar{y}) - T\left(\bar{y}+\frac{\bar{y}}{x_i}\right)\right] \quad (12)$$

where $T(z) = e^z\text{Ei}\left(-z\right)$ and $\text{Ei(z)} = -\int_{-z}^{\infty} e^{-t}\; t^{-1}dt$ is the exponential integration [15]. In (12), $x_i = \sigma_n^2/aP_i(g_{ss})$, $\bar{y} = \sigma_n^2/\bar{\sigma}_p^2$ and $\bar{\sigma}_p^2 = \mathbb{E}_{g_{ps}}\{\sigma_p^2\} = P_p\gamma_{ps}L_{ps}A(\phi_r - \theta_p')$. Finally, the ergodic capacity in this scenario is

$$C = \mathbb{E}_{g_{ss}}\left\{\sum_{i=0}^{1}\left[\hat{\pi}_i\log_2\left(1+\frac{1}{x_i}\right) + \frac{\beta_i}{\ln(2)}\Big(T(\bar{y}) - T\big(\bar{y}+\frac{\bar{y}}{x_i}\big)\Big)\right]\right\}$$

Moreover, the constraints in (4) and (5) can be written as

$$\mathbb{E}_{g_{ss}}\{\bar{b}_0 P_0(g_{ss}) + \bar{b}_1 P_1(g_{ss})\} \leq \bar{I}_{\text{av}} \quad (13a)$$

$$\mathbb{E}_{g_{ss}}\{\hat{\pi}_0 P_0(g_{ss}) + \hat{\pi}_1 P_1(g_{ss})\} \leq \bar{P}_{\text{av}} \quad (13b)$$

where $\bar{b}_i = \mathbb{E}_{g_{sp}}\{b_i\} = \beta_i\gamma_{sp}L_{sp}A(\phi_t - \theta_p)$. The optimal transmit power levels $P_i(g_{ss})$ can be obtained by solving the following equation

$$\frac{\partial L}{\partial P_i(g_{ss})} = \frac{-a}{\sigma_n^2\ln(2)}f_i(x_i,\bar{y}) + \lambda\hat{\pi}_i + \mu\bar{b}_i = 0$$

where

$$f_i(x,\bar{y}) = \frac{\alpha_i x}{x+1} - \beta_i\bar{y}\; T\big(\bar{y}+\frac{\bar{y}}{x}\big).$$

This equation has no closed form solution and has to be solved numerically. Furthermore, the parameter $Z$ in (11) for this scenario is modified to

$$\bar{Z} = \frac{\bar{I}_{\text{av}}}{\pi_1 \gamma_{sp} A_0 \bar{P}_{\text{av}}} - \frac{A_1}{A_0}. \tag{14}$$

Algorithm 1 can be used for this scenario with some modifications.

### C. Statistical CSI for All Fading Channels

In the third scenario we assume that $\text{SU}_{\text{tx}}$ cannot estimate $g_{ss}$ and it knows only the statistical CSI of all fading channels. Even if $\text{SU}_{\text{tx}}$ can estimate $g_{ss}$, when we maximize the capacity for each realization of $g_{ss}$, the optimal $\phi_t$ and $\phi_r$ will be a function of $g_{ss}$ and as a result they may change very fast in a fast fading environment. In some cases where antennas are steered mechanically, their rotation speeds are limited and cannot adapt themselves according to channel variations. Thus, in this scenario we wish the optimal directions to be independent of the realizations of fading coefficients. Hence, we take expectation with respect to all fading coefficients and then maximize capacity. The expectation of $c_{0,i}$ is equal to $\mathbb{E}_{g_{ss}}\{c_{0,i}\} = -T(\bar{x}_i)/\ln(2)$, where $\bar{x}_i = \sigma_n^2/\bar{a}P_i$ and $\bar{a} = \mathbb{E}\{a\} = \gamma_{ss} L_{ss} G$. Similar to previous section, we can write $\mathbb{E}_{\boldsymbol{g}}\{c_{1,i}\} = -U(\bar{x}_i, \bar{y})/\ln(2)$ where

$$U(\bar{x}_i, \bar{y}) = \begin{cases} \frac{-\bar{y}}{\bar{y}-\bar{x}_i}\left[T(\bar{y}) - T(\bar{x}_i)\right], & \text{if } \bar{x}_i \neq \bar{y} \\ -\bar{x}_i T(\bar{x}_i) - 1. & \text{if } \bar{x}_i = \bar{y} \end{cases}$$

The ergodic capacity is

$$C = \frac{-1}{\ln(2)} \sum_{i=0}^{1} \left[\alpha_i T(\bar{x}_i) + \beta_i U(\bar{x}_i, \bar{y})\right]$$

and the constraints in (4) and (5) can be written as

$$\bar{b}_0 P_0 + \bar{b}_1 P_1 \leq \bar{I}_{\text{av}} \tag{15a}$$

$$\hat{\pi}_0 P_0 + \hat{\pi}_1 P_1 \leq \bar{P}_{\text{av}}. \tag{15b}$$

The optimal transmit power levels can be obtained by solving the following equation numerically

$$\frac{\partial L}{\partial P_i} = \frac{-\bar{a}}{\sigma_n^2 \ln(2)} h_i(\bar{x}_i, \bar{y}) + \lambda \hat{\pi}_i + \mu \bar{b}_i = 0$$

where

$$h_i(\bar{x}, \bar{y}) = \bar{x}^2 \left(\alpha_i \frac{\partial T(\bar{x})}{\partial x} + \beta_i \frac{\partial U(\bar{x}, \bar{y})}{\partial x}\right).$$

Algorithm 1 can be used for this scenario.

## IV. Numerical Results

We numerically show the effect of using directional antennas on the ergodic capacity of the considered CR system when spectrum sensing is imperfect. Assume $\sigma_n^2 = 1$, $\phi_{3\text{dB}} = 45°$, $A_0 = 9.8$, $A_1 = 0.2$, $\gamma_{ss} = \gamma_{sp} = \gamma_{ps} = 1$, $\pi_1 = 0.3$, $\theta_p = 90°$ and $\theta'_p = 130°$. For fair comparisons, we consider a fixed spectrum sensing method with $P_d = 0.9$ and $P_f = 0.1$.

Suppose $C_{\text{opt}}^{\text{Dir}}$ denote the optimal capacity when we use directional antennas. Fig. 3 shows $C_{\text{opt}}^{\text{Dir}}$ versus $\theta$ for $P_p = 0.4, 3$ watts for all three scenarios when $\bar{P}_{\text{av}} = 12$ dB. When $\theta$ increases from 0° to 40°, $\text{SU}_{\text{rx}}$ receives less interference from $\text{PU}_{\text{tx}}$ and $\text{SU}_{\text{tx}}$ can increase transmit power and as a result the capacity increases. However, when $\theta$ increases from 40° to 80°, $\text{SU}_{\text{tx}}$ imposes more interference on $\text{PU}_{\text{rx}}$ and the optimal capacity decreases. Furthermore, we observe that the capacity for scenario 3 is always smaller than that of scenarios 1 and 2. Increasing $P_p$ doesn't have any impact on constraints, however, the capacity expression depends on $P_p$ and as it can be seen in Fig. 3, increasing $P_p$ decreases the capacity. Fig. 4 shows the optimal capacity for all three scenarios when $\bar{P}_{\text{av}} = 15$ dB. Comparing Figs. 3 and 4, we can see that when the maximum allowed average transmit power of $\text{SU}_{\text{tx}}$ ($\bar{P}_{\text{av}}$) increases, the capacity increases as well, provided that the constraint (4) is not violated. Fig. 5 which plots $C_{\text{opt}}^{\text{Dir}}$ versus $\bar{P}_{\text{av}}$ when $\theta = 50°$ and $\bar{I}_{\text{av}} = 0$ dB also shows the similar fact.

Let $C_{\text{opt}}^{\text{Omn}}$ denote the capacity when $\text{SU}_{\text{tx}}$ and $\text{SU}_{\text{rx}}$ have omni-directional antennas and only transmit power levels $P_0$ and $P_1$ are optimized subject to constraints (4) and (5). Note that $P_0^{\text{opt}}$ and $P_1^{\text{opt}}$ are constant for all $\theta$ when SUs use omni-directional antennas and $C_{\text{opt}}^{\text{Omn}}$ is independent of $\theta$. Furthermore, let $C_{\text{opt}}^{\text{LOS}}$ be the capacity when directional antennas of $\text{SU}_{\text{tx}}$ and $\text{SU}_{\text{rx}}$ are exactly pointed at each other ($\phi_t = \theta$, $\phi_r = \pi + \theta$) and only $P_0$ and $P_1$ are optimized subject to constraints (4) and (5). We compare $C_{\text{opt}}^{\text{Dir}}$, $C_{\text{opt}}^{\text{Omn}}$ and $C_{\text{opt}}^{\text{LOS}}$.

We define three capacity ratios $\Gamma_{\text{D2O}} = C_{\text{opt}}^{\text{Dir}}/C_{\text{opt}}^{\text{Omn}}$, $\Gamma_{\text{L2O}} = C_{\text{opt}}^{\text{LOS}}/C_{\text{opt}}^{\text{Omn}}$ and $\Gamma_{\text{D2L}} = C_{\text{opt}}^{\text{Dir}}/C_{\text{opt}}^{\text{LOS}}$. Fig. 6 plots $\Gamma_{\text{D2O}}$ and $\Gamma_{\text{L2O}}$ versus $\theta$ when $\bar{P}_{\text{av}} = 12, 15$ dB. We observe that when $\theta \approx \theta_p$, $C_{\text{opt}}^{\text{Dir}} \approx C_{\text{opt}}^{\text{LOS}}$ and as $|\theta - \theta_p|$ increases, the capacity gain increases. When $\text{PU}_{\text{rx}}$ and $\text{SU}_{\text{rx}}$ are close, using directional antennas does not enhance the ergodic capacity (with respect to using omni-directional antennas). The capacity gain in Fig. 6 finally saturates, since the direction of $\text{SU}_{\text{tx}}$ goes sufficiently away from $\text{PU}_{\text{rx}}$ and directional antenna of $\text{SU}_{\text{tx}}$ reduces the interference imposed on $\text{PU}_{\text{rx}}$. In addition, we can see that when $\bar{P}_{\rm av}$ of $\text{SU}_{\text{tx}}$ increases , the ergodic capacity increases, while constraints (4) and (5) still hold true.

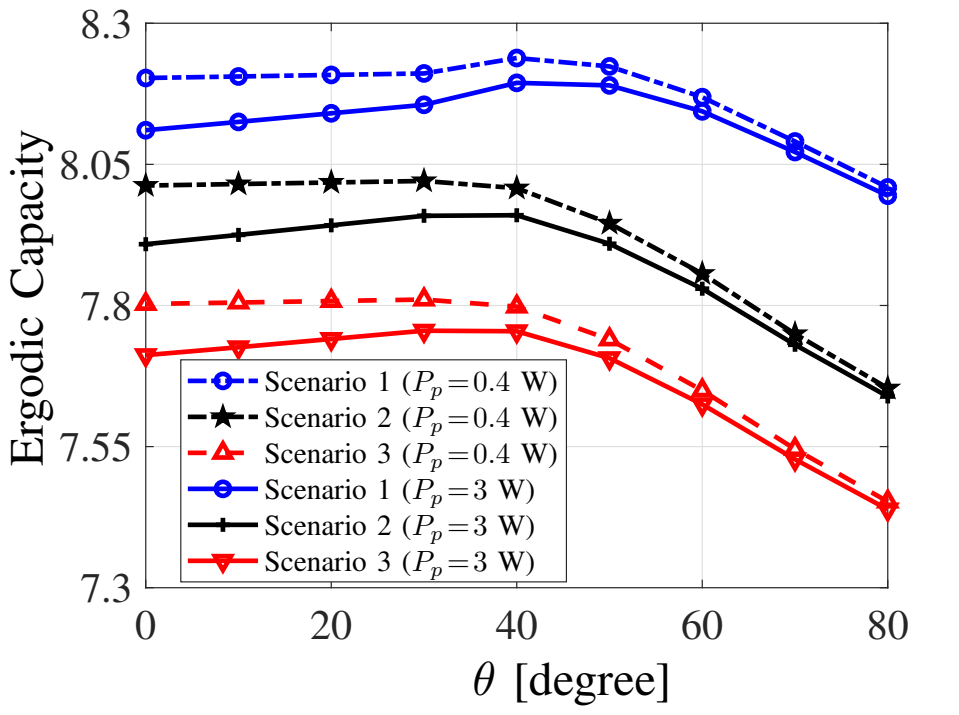

Fig. 3: $C_{\text{opt}}^{\text{Dir}}$ versus $\theta$ for three scenarios when $\bar{P}_{\text{av}} = 12$ dB.

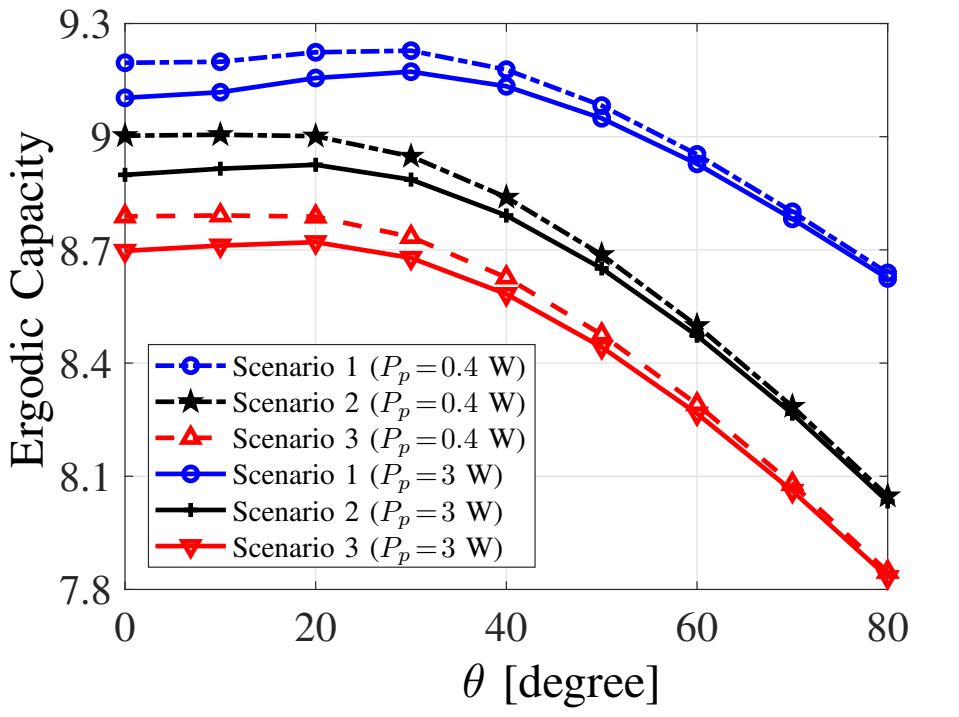

Fig. 4: $C_{\text{opt}}^{\text{Dir}}$ versus $\theta$ for three scenarios when $\bar{P}_{\text{av}} = 15$ dB.

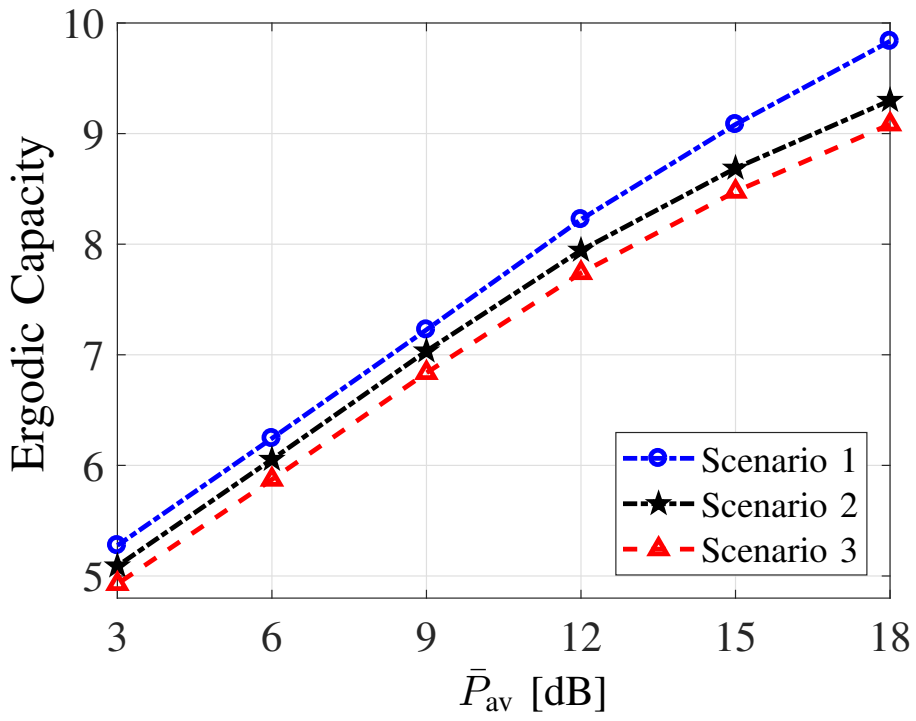

Fig. 5: $C_{\rm opt}^{\rm Dir}$ versus $\bar{P}_{\rm av}$ for three scenarios when $\theta = 50°$, $\bar{I}_{\rm av} = 0$.

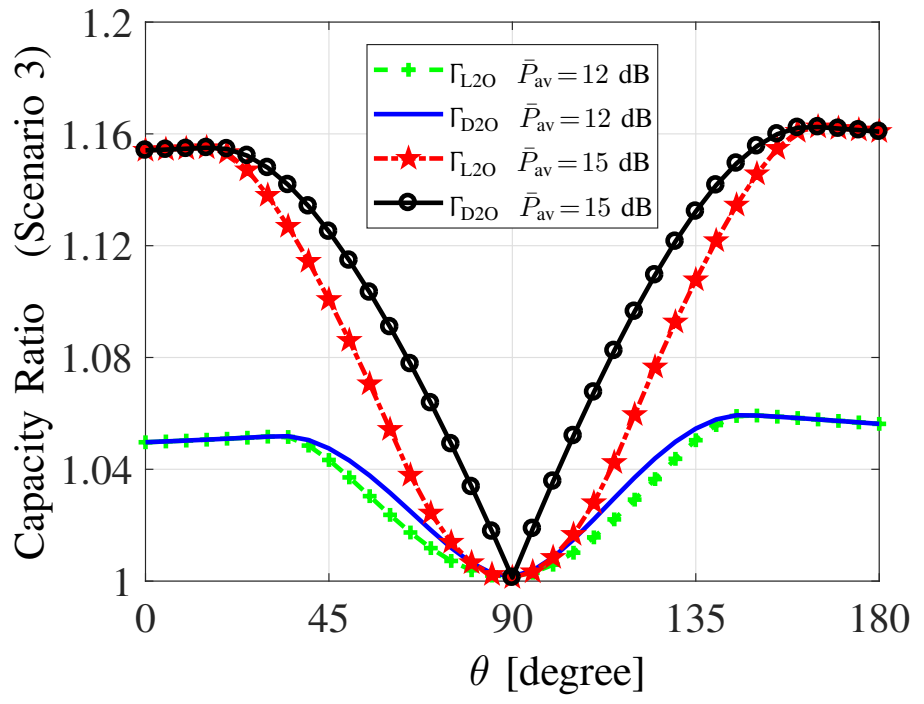

Fig. 6: Capacity ratio versus $\theta$ when $\bar{I}_{\rm av} = 0$ dB.

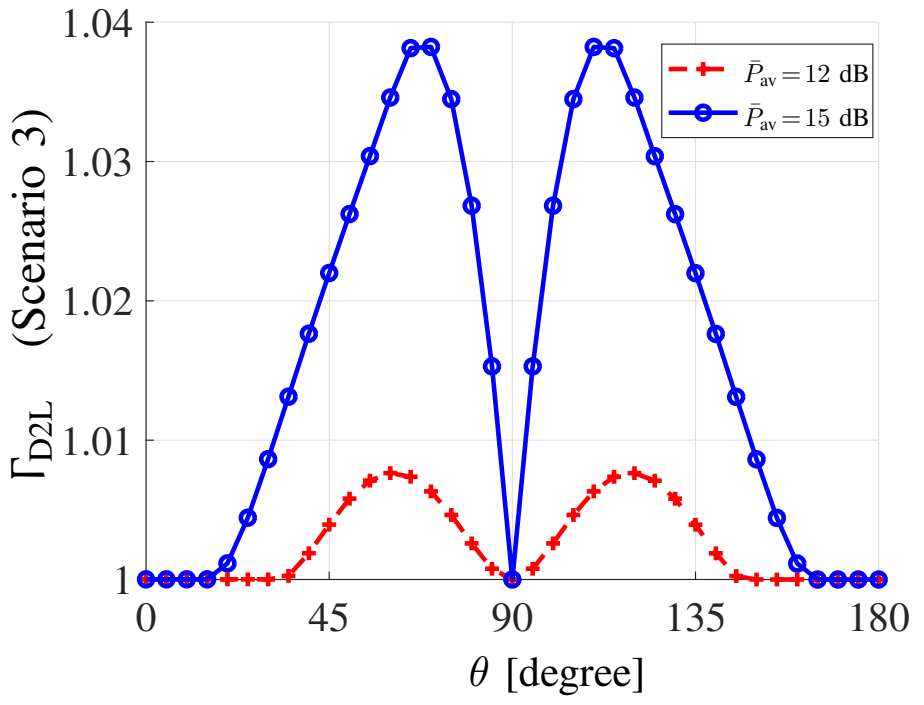

Fig. 7: $\Gamma_{\rm D2L}$ versus $\theta$ when $\bar{I}_{\rm av} = 0$ dB.

The effect of optimizing the orientation of directional antennas on ergodic capacity is illustrated in Fig. 7, where the capacity gain $\Gamma_{\rm D2L}$ versus $\theta$ is plotted for $\bar{P}_{\rm av} = 12, 15$ dB. We note that when we optimize the angles $\phi_t$ and $\phi_r$, $\text{SU}_{\text{tx}}$ can use more power for transmission (i.e., use higher power levels $P_0$ and $P_1$) without violating constraints (4) and (5) and, hence, the capacity increases.

## V. Conclusion

In this paper, we considered a CR system, where the SUs are equipped with steerable directional antennas. The $\text{SU}_{\text{tx}}$ first senses the spectrum (with error) and then transmits data at two power levels, according to the result of sensing. The optimal $\text{SU}_{\text{tx}}$ transmit power levels and the optimal directions of $\text{SU}_{\text{tx}}$ transmit antenna and $\text{SU}_{\text{rx}}$ receive antenna are obtained by maximizing the ergodic capacity, subject to average transmit power and average interference power constraints. To study the effect of fading channels, we considered three scenarios: 1) when $\text{SU}_{\text{tx}}$ knows fading channels between $\text{SU}_{\text{tx}}$ and $\text{PU}_{\text{rx}}$, $\text{PU}_{\text{tx}}$ and $\text{SU}_{\text{rx}}$, $\text{SU}_{\text{tx}}$ and $\text{SU}_{\text{rx}}$, 2) when $\text{SU}_{\text{tx}}$ knows only the channel between $\text{SU}_{\text{tx}}$ and $\text{SU}_{\text{rx}}$, and statistics of the other two channels, and, 3) when $\text{SU}_{\text{tx}}$ only knows the statistics of these three fading channels. Through simulations, we showed that directional antennas significantly enhance the ergodic capacity, without violating the power constraints.

## Acknowledgment

This research is supported by NSF under grant ECCS-1443942.